\documentclass[aps,prd,twocolumn,superscriptaddress,nofootinbib,10pt,secnumarabic]{revtex4-1}

\usepackage[colorlinks,linkcolor=blue,anchorcolor=blue,citecolor=blue,urlcolor=blue]{hyperref}
\usepackage{graphicx,subfigure}
\usepackage[figuresright]{rotating}
\usepackage{amsmath,amsfonts,amssymb,bm}
\usepackage{array,enumitem,multirow}
\usepackage{acronym}
\usepackage{makecell}
\usepackage{orcidlink}
\usepackage{xcolor}

\newcommand{\be}{\begin{equation}}
\newcommand{\ee}{\end{equation}}
\newcommand{\bea}{\begin{eqnarray}}
\newcommand{\eea}{\end{eqnarray}}

\begin{document}

\title{GW231123: False Massive Graviton Signatures from Unmodeled Point-Mass Lensing}

\author{Baoxiang Wang\, \orcidlink{0009-0003-3998-4609}}
\affiliation{School of Physics and Technology, Wuhan University, Wuhan 430072, China}

\author{Tao Yang\, \orcidlink{0000-0002-2161-0495}}
\thanks{Corresponding author: yangtao@whu.edu.cn}
\affiliation{School of Physics and Technology, Wuhan University, Wuhan 430072, China}

\date{\today}

\begin{abstract}
GW231123 is the strongest current candidate for a lensed gravitational wave event and a unique case for testing how point-mass lensing affects propagation-based tests of gravity. In the real GW231123 data, an unlensed IMRPhenomXPHM analysis yields an apparent nonzero graviton mass posterior. We show that this anomaly is naturally explained by unmodeled point-mass lensing: once lensing is included, the apparent graviton mass signal disappears. In GW231123-like injection-recovery tests, a lensed NRSur7dq4 signal with zero graviton mass, recovered with the same unlensed IMRPhenomXPHM template, produces a similarly pronounced spurious graviton mass posterior, whereas lensing-included analyses with IMRPhenomXPHM, IMRPhenomXO4a, and NRSur7dq4 remain mutually consistent with no evidence for nonzero graviton mass. The similarity between the injected and real data posteriors shows that unmodeled point-mass lensing can mimic modified gravitational wave propagation. These results identify a concrete failure mode in tests of gravity and strengthen the interpretation of GW231123 as a lensed candidate.
\end{abstract}

\maketitle

\section{Introduction}
After roughly a decade of gravitational wave observations and the analysis of more than 200 events, no lensed gravitational wave signal has yet been established conclusively \cite{Hannuksela:2019kle, LIGOScientific:2021izm, LIGOScientific:2023bwz, Janquart:2023mvf, Chakraborty:2025maj, LIGOScientific:2025cwb, Haris:2018vmn, Li:2019osa, Janquart:2021qov, Goyal:2021hxv}. GW231123 therefore occupies a special place in the current catalog. In the GWTC-4.0 lensing searches, it is the only event identified as an outlier under an isolated point-mass lens model, although waveform inaccuracies or other unmodeled features could still artificially enhance the support for lensing \cite{LIGOScientific:2025cwb}. GW231123 is thus not a confirmed lensed signal, but it remains the strongest current candidate for a lensed gravitational wave event \cite{LIGOScientific:2025cwb}.

This possible lensing interpretation is especially consequential because GW231123 is also anomalous in tests of gravitational wave propagation. Modified-dispersion searches provide a standard framework for constraining departures from general relativity, including an effective graviton mass \cite{Will1998,Mirshekari2012,Abbott2016GW150914GR,GWTC1TGR,GWTC2TGR,GWTC3TGR}. Although the GWTC-4.0 parameterized tests remain broadly consistent with general relativity at the catalog level \cite{LIGOScientific:2026tgr2}, that analysis reported a substantial discrepancy for GW231123 between IMRPhenomXPHM \cite{Pratten:2020ceb} and NRSur7dq4 \cite{Varma:2019csw}. In the graviton mass case, the corresponding Bayes factors are $\log_{10} B^{\rm MDR}_{\rm GR}=1.0$ for IMRPhenomXPHM and $-1.3$ for NRSur7dq4. GW231123 is therefore excluded from the combined catalog bound on gravitational wave dispersion because of waveform systematic uncertainties \cite{LIGOScientific:2026tgr2}.

These two anomalies suggest a common origin. If GW231123 is affected by point-mass wave-optics lensing, then an unlensed waveform family may absorb the missing frequency-dependent lensing distortion into a propagation parameter and thereby produce a false massive-graviton signature. In this Letter, we test this possibility directly using both the real GW231123 data and controlled injection-recovery studies. Using an unlensed IMRPhenomXPHM template augmented by a massive-graviton propagation term, we find an apparent nonzero graviton-mass posterior. Once point-mass lensing is included, however, this preference disappears. Moreover, with lensing included, the inferences obtained with IMRPhenomXPHM, IMRPhenomXO4a \cite{Thompson:2023ase}, and NRSur7dq4 become mutually consistent and show no evidence for nonzero graviton mass. This indicates that the apparent graviton-mass signal is more naturally explained by unmodeled lensing than by modified propagation.

We then turn this interpretation into a controlled test. We construct a GW231123-like injection using the maximum-posterior parameters from a lensed NRSur7dq4 analysis and inject a point-mass-lensed signal with $m_g=0$. The choice of NRSur7dq4 is deliberate: in the lensing analysis it is described as the most accurate waveform model on average in this region of parameter space, and in the GWTC-4.0 modified-dispersion study it serves as the key cross-check against waveform systematics \cite{LIGOScientific:2025cwb,LIGOScientific:2026tgr2}. Recovering this signal with the same unlensed IMRPhenomXPHM template used in the real-data analysis reproduces the key anomaly: although the injected signal has $m_g=0$, the recovery yields a pronounced apparent nonzero graviton-mass posterior that is quantitatively similar to that obtained from the real GW231123 data. By contrast, analyses that include point-mass lensing remain consistent with no evidence for nonzero graviton mass, in agreement with the behavior already seen in the real event.

GW231123 is therefore more than an isolated outlier in lensing searches. It is the only current event in which candidate lensing, strong waveform dependence, and an apparent graviton-mass anomaly appear together in a single system. Our real-data and injection-recovery results show that these features can be understood within a single mechanism: a point-mass-lensed signal analyzed with an unlensed waveform family can produce a false graviton-mass signature. In this sense, our results not only expose a concrete failure mode in propagation-based tests of gravity, but also strengthen the interpretation of GW231123 as a lensed candidate. More broadly, they show that tests of general relativity with candidate lensed events must model point-mass wave-optics distortions explicitly; otherwise, template incompleteness may be misidentified as evidence for new gravitational physics.

\section{Point-mass microlensing with massive graviton dispersion}
\label{sec:microlensing_mg}
We begin by incorporating the effect of a nonzero graviton mass as a propagation-induced dispersion in the frequency-domain waveform. Following the standard phenomenological treatment \cite{Will1998,Mirshekari2012,GWTC1TGR,GWTC2TGR,GWTC3TGR}, if the graviton has rest mass $m_g$, then the GW group velocity satisfies
\begin{equation}
    \frac{v_g^2}{c^2}
    =
    1-\frac{m_g^2 c^4}{E^2},
    \label{eq:vg_exact}
\end{equation}
where $E$ is the graviton energy. In the relativistic regime relevant for ground-based GW observations, this becomes
\begin{equation}
    v_g(f)
    \simeq
    c\left[1-\frac{1}{2}\left(\frac{c}{\lambda_g f}\right)^2\right],
    \label{eq:vg_mg}
\end{equation}
where $\lambda_g = h/(m_g c)$ is the graviton Compton wavelength. A finite graviton mass therefore causes lower-frequency Fourier components to propagate slightly more slowly than higher-frequency ones, producing an accumulated dephasing during propagation \cite{Will1998,Mirshekari2012}.

Let $h_{\rm UL}(f)$ denote the unlensed frequency-domain waveform. massive graviton propagation then enters as an additional dispersive phase,
\begin{equation}
    h_{\rm disp}(f;m_g)
    =
    h_{\rm UL}(f)\,e^{i\Psi_{\rm disp}(f;m_g)},
    \label{eq:disp_waveform}
\end{equation}
with
\begin{equation}
    \Psi_{\rm disp}(f;m_g)
    =
    -\,\frac{\pi D_0}{\lambda_g^2(1+z)\,f},
    \label{eq:disp_phase}
\end{equation}
where $z$ is the source redshift and $D_0$ is the effective cosmological distance factor that enters the massive graviton phase correction. In the limit of vanishing graviton mass, or equivalently infinite graviton Compton wavelength, the dispersive phase vanishes and the standard general-relativistic waveform is recovered.

We now include point-mass microlensing in the wave-optics regime, which modifies the observed GW signal through a frequency-dependent amplification factor \cite{TakahashiNakamura2003,DaiVenumadhav2017,ChristianVitaleLoeb2018,JungEtAl2019,Diego2019,Pagano2020,GrespanBiesiada2023, Nakamura:1997sw, Nakamura:1999uwi, Takahashi:2003ix}. In the presence of a compact intervening lens, the full lensed and dispersive waveform is written as \cite{ChungLi2021}
\begin{equation}
    h_{\rm ML}(f;M_L^z,y,m_g)
    =
    F(f;M_L^z,y,m_g)\,
    h_{\rm UL}(f)\,
    e^{i\Psi_{\rm disp}(f;m_g)},
    \label{eq:full_waveform_mg}
\end{equation}
where $F(f;M_L^z,y,m_g)$ is the complex amplification factor, $M_L^z \equiv M_L(1+z_L)$ is the redshifted lens mass, and $y$ is the dimensionless impact parameter in units of the Einstein radius.

For an isolated point-mass lens, the standard wave-optics amplification factor depends on the dimensionless diffraction parameter $w = 8\pi G M_L^z f/c^3$ together with the impact parameter $y$. In the presence of massive graviton dispersion, the relevant frequency variable in the diffraction kernel is modified by \cite{ChungLi2021}
\begin{equation}
    \beta(f)\equiv \frac{c}{v_g(f)}
    \simeq 1+\frac{1}{2}\left(\frac{c}{\lambda_g f}\right)^2.
    \label{eq:beta_mg}
\end{equation}
Equivalently, the point-mass amplification factor is obtained from the standard expression by replacing the diffraction parameter $w$ with $w\,\beta(f)$. Following standard notation \cite{TakahashiNakamura2003,ChungLi2021}, this gives
\begin{equation}
\begin{aligned}
F(f;M_L^z,y,m_g)
  ={}& \exp\!\Biggl[
        \frac{\pi w\beta}{4}
        + \frac{i w\beta}{2}\ln\!\left(\frac{w\beta}{2}\right)
        - i w\beta\,\phi_m(y)
      \Biggr]
\\
  &\times \Gamma\!\left(1-\frac{i w\beta}{2}\right)
   \times {}_1F_1\!\left(
          \frac{i w\beta}{2},
          1;\,
          \frac{i w\beta y^2}{2}
       \right),
\end{aligned}
\label{eq:ampfactor_mg}
\end{equation}
where $\phi_m(y) = (x_m-y)^2/2-\ln(x_m)$ and $x_m = (y+\sqrt{y^2+4})/2$. Here $M_L^z$ sets the characteristic frequency scale of the wave-optics modulation, while $y$ controls the strength of the interference pattern: small $y$ produces stronger oscillatory distortions, whereas large $y$ approaches the unlensed limit, with $F$ approaching unity.

Equation~(\ref{eq:full_waveform_mg}) defines the waveform model used throughout this work. We implement the point-mass microlensing modification in the frequency domain using the publicly available \texttt{GWMAT} package \cite{Mishra:GWMAT}, which augments standard BBH waveform models with the lens parameters $(M_L^z, y)$ and applies the corresponding amplification factor to the unlensed waveform. In the limit of vanishing graviton mass, it reduces to the standard lensed waveform in general relativity; in the absence of lensing, it reduces to the usual dispersive waveform with massive graviton propagation.

For GW231123, this combined treatment is essential. Because only a few cycles are observed in band, even modest frequency-dependent distortions can bias the inferred parameters if they are not modeled consistently \cite{LIGOScientific:2025rsn,LIGOScientific:2025cwb,Mishra2024MicrolensingImpact,Mishra2024TGRBias}. Point-mass microlensing and massive graviton dispersion therefore need to be treated jointly when deriving graviton mass constraints from a candidate lensed event.

Unless otherwise stated, we adopt the weakly informative priors commonly used in point-mass microlensing searches. We assume a log-uniform prior on the redshifted lens mass, $p(\log_{10} M_L^z)\propto {\rm Uniform}(0,5)$, and a prior $p(y)\propto y$ for $y\!\in\!(0.1,3)$, motivated by simple geometric arguments for randomly distributed source-lens alignments. When posteriors approach the prior boundaries, we enlarge the corresponding ranges to avoid artificial truncation. These choices provide broad coverage of the microlensing parameter space while reducing sensitivity to prior-volume effects \cite{Pagano2020,LIGOScientific:2025cwb,Mishra2024MicrolensingImpact}.

\section{Bayesian Analysis}
\label{sec:bayes}
We analyze GW231123 within the standard Bayesian framework, treating each waveform model, defined by specific assumptions about massive graviton propagation and point-mass microlensing, as a separate hypothesis $H$ \cite{Skilling2004,Biwer2019,Speagle2020,Ashton2019Bilby,Narola:2023men,Chan:2025pdf}. For a given hypothesis and detector data $d$, the posterior distribution for the parameter vector $\boldsymbol{\theta}$ is obtained from Bayes' theorem,
\begin{equation}
p(\boldsymbol{\theta}\,|\,d,H)
 = \frac{\pi(\boldsymbol{\theta}\,|\,H)\,
         \mathcal{L}(d\,|\,\boldsymbol{\theta},H)}
        {Z_H},
\end{equation}
where $\pi(\boldsymbol{\theta}\,|\,H)$ denotes the prior, $\mathcal{L}(d\,|\,\boldsymbol{\theta},H)$ the likelihood, and $Z_H=\int \mathrm{d}\boldsymbol{\theta}\,\pi(\boldsymbol{\theta}\,|\,H)\,\mathcal{L}(d\,|\,\boldsymbol{\theta},H)$ is the Bayesian evidence for hypothesis $H$.

For GW231123, we use strain data from the LIGO Hanford and Livingston observatories. The likelihood is evaluated on a 16 s data segment centered on the trigger time, and only frequencies above 20 Hz are included in the analysis, matching the configuration adopted in the original LVK study of this event \cite{LIGOScientific:2025rsn}. The power spectral densities used for both detectors are also identical to those employed in that analysis, ensuring that any differences in inference arise from the waveform physics rather than from differences in data conditioning \cite{LIGOScientific:2025rsn,GWOSC2020}.

Assuming stationary, Gaussian, and uncorrelated noise, the log-likelihood for parameters $\boldsymbol{\theta}$ takes the form
\begin{equation}
\ln \mathcal{L}(d\,|\,\boldsymbol{\theta},H)
 = -\frac{1}{2}
   \sum_k
   \big\langle d_k-h_k(\boldsymbol{\theta})
     \,\big|\,
     d_k-h_k(\boldsymbol{\theta})
   \big\rangle
   + \mathrm{const},
\end{equation}
with noise-weighted inner product
\begin{equation}
\big\langle a \,\big|\, b \big\rangle
 = 2 \int_{f_{\rm low}}^{f_{\rm high}}
   \frac{\tilde{a}^*(f)\tilde{b}(f)
        + \tilde{a}(f)\tilde{b}^*(f)}
        {S_n(f)}\,\mathrm{d}f .
\end{equation}

The parameter vector $\boldsymbol{\theta}$ includes the standard binary-black-hole parameters together with the additional quantities required by the extended waveform models considered here. For models including massive graviton propagation, the parameter set is augmented by the graviton mass parameter $m_g$, or equivalently the graviton Compton wavelength $\lambda_g$ \cite{Will1998,Mirshekari2012}. For models including microlensing, it is further augmented by the redshifted lens mass $M_L^z$ and the dimensionless impact parameter $y$. We adopt the same priors as in the LVK analysis of GW231123 for all standard source parameters, together with weakly informative priors on $(M_L^z,y)$ for lensing hypotheses \cite{LIGOScientific:2025rsn,LIGOScientific:2025cwb,Pagano2020}.

Posterior sampling and evidence evaluation are performed using \textsc{PyCBC} with the \textsc{Dynesty} nested-sampling algorithm \cite{Biwer2019,Speagle2020}, which simultaneously explores the posterior distribution and evaluates the multidimensional integral defining $Z_H$. This setup provides consistent estimates of both posterior densities and evidences across all waveform hypotheses, and the sampling uncertainty in $\ln Z_H$ is propagated directly into the quoted Bayes factors.

To determine which waveform description is more strongly supported by the data, we compare hypotheses using the Bayes factor 

\begin{equation}
B^{H_2}_{H_1}=\frac{Z_{H_2}}{Z_{H_1}}. 
\end{equation}

Throughout this work, Bayes factors are reported in base-10 logarithmic form, $\log_{10} B^{H_2}_{H_1}=\log_{10}(Z_{H_2}/Z_{H_1})$. The hypotheses considered here are constructed so that both include the massive graviton parameter, while differing only in whether point-mass microlensing is incorporated in the waveform model. In this way, comparisons between lensed and unlensed models isolate the impact of microlensing on the inference of the graviton mass.

\section{Results}
\label{sec:results}
We begin with the real GW231123 data and analyze the event with the \texttt{IMRPhenomXPHM} waveform including a massive graviton propagation term, comparing the results obtained with and without point-mass microlensing. In our sampling setup, the inference is performed using a prior uniform in $\log_{10}\lambda_g$; we then transform the posterior samples to $m_g$ and reweight them to the corresponding posterior under a prior uniform in $m_g$.

\begin{figure*}[t]
    \centering
    \includegraphics[width=\textwidth]{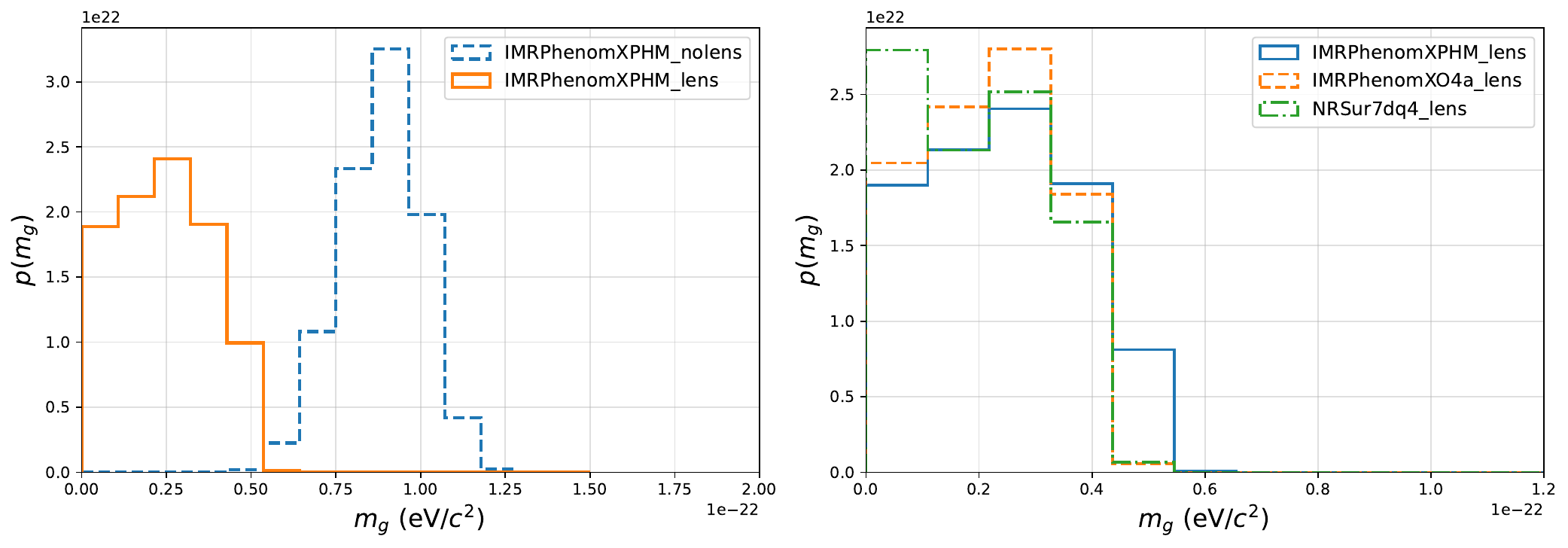}
    \caption{
    Left: unlensed and lensed \texttt{IMRPhenomXPHM} posteriors for the real GW231123 data.
    Right: lensed posteriors for the real GW231123 data obtained with \texttt{IMRPhenomXPHM}, \texttt{IMRPhenomXO4a}, and \texttt{NRSur7dq4}.
    The apparent support for nonzero graviton mass in the unlensed analysis disappears once point-mass lensing is included, and the lensing-included results become mutually consistent across waveform models.
    }
    \label{fig:real_two_panel}
\end{figure*}

The results for the real event are summarized in Fig.~\ref{fig:real_two_panel}. The left panel compares the graviton mass posteriors inferred with \texttt{IMRPhenomXPHM} under the unlensed and lensed hypotheses. Under the unlensed hypothesis, the posterior exhibits an apparent preference for nonzero graviton mass. Once the waveform is extended to include the point-mass lens amplification factor, this preference disappears and the posterior becomes consistent with the massless limit.

We quantify this effect by comparing the lensed and unlensed hypotheses with the corresponding Bayes factor while retaining the massive graviton parameter in both models. For \texttt{IMRPhenomXPHM}, we find $\log_{10} B^{\rm lens}_{\rm nolens}=3.6$, showing that the data strongly favor the lensing-included hypothesis.

The right panel of Fig.~\ref{fig:real_two_panel} tests whether this conclusion remains stable across waveform families. We compare the graviton mass posteriors obtained with \texttt{IMRPhenomXPHM}, \texttt{IMRPhenomXO4a}, and \texttt{NRSur7dq4}, all including point-mass microlensing. Once lensing is incorporated, the three waveform models become mutually consistent and show no evidence for nonzero graviton mass.

We next test whether this behavior can arise directly from unmodeled lensing by performing GW231123-like injection-recovery experiments. We inject a lensed \texttt{NRSur7dq4} signal with $m_g=0$, using the maximum-posterior parameters obtained from the lensed GW231123 analysis. The choice of \texttt{NRSur7dq4} follows its role as the key cross check against waveform systematics in previous analyses.

\begin{figure*}[t]
    \centering
    \includegraphics[width=\textwidth]{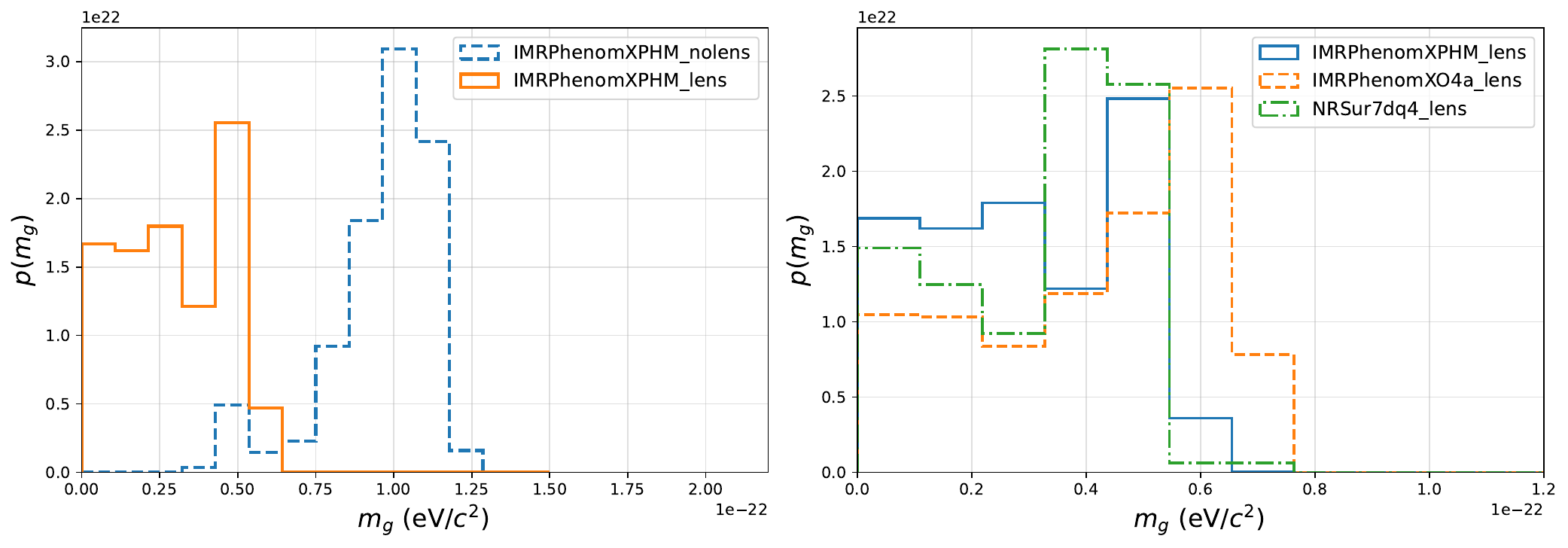}
    \caption{
    Left: unlensed and lensed \texttt{IMRPhenomXPHM} posteriors for the GW231123-like injection.
    Right: lensed posteriors for the same injection obtained with \texttt{IMRPhenomXPHM}, \texttt{IMRPhenomXO4a}, and \texttt{NRSur7dq4}.
    For the lensed \texttt{NRSur7dq4} injection with $m_g=0$, the unlensed recovery produces a spurious apparent preference for nonzero graviton mass, whereas the lensing-included recoveries remain mutually consistent.
    }
    \label{fig:inj_two_panel}
\end{figure*}

The injection results are summarized in Fig.~\ref{fig:inj_two_panel}. The left panel compares the recovery of the injected signal with the same \texttt{IMRPhenomXPHM} waveform used in the real-data analysis, again under unlensed and lensed hypotheses. The key anomaly is reproduced: although the injected signal has $m_g=0$, the unlensed recovery yields a pronounced apparent preference for nonzero graviton mass, while the lensed recovery remains consistent with the massless limit. The resulting unlensed posterior closely resembles that obtained from the real GW231123 data with the same waveform model. Since the injection contains no modified propagation, this shows that the apparent massive graviton signal in GW231123 can arise from recovering a lensed general-relativistic signal with an unlensed waveform family.

The right panel of Fig.~\ref{fig:inj_two_panel} shows the corresponding lensed recoveries with \texttt{IMRPhenomXPHM}, \texttt{IMRPhenomXO4a}, and \texttt{NRSur7dq4}. As in the real-data analysis, the posteriors obtained with the three lensing-included waveform models remain mutually consistent and show no evidence for nonzero graviton mass.

Taken together, the real data and injection recovery results support a common interpretation: the apparent graviton mass anomaly in GW231123 is best understood as the consequence of unmodeled point-mass microlensing. Adopting the lensing-included interpretation, we find no evidence for nonzero graviton mass and obtain a 90\% credible upper bound of $m_g < 4.3\times10^{-23}\,\mathrm{eV}/c^2$.

\section{Conclusion and Discussion}
GW231123 is the strongest current candidate for a lensed gravitational wave event and a unique system for testing how point-mass lensing affects propagation-based tests of gravity. In the real data, an unlensed \texttt{IMRPhenomXPHM} analysis yields an apparent nonzero graviton-mass posterior, but this preference disappears once point-mass lensing is included. With lensing included, the inferences obtained with \texttt{IMRPhenomXPHM}, \texttt{IMRPhenomXO4a}, and \texttt{NRSur7dq4} become mutually consistent and show no evidence for nonzero graviton mass.

The key new result comes from the GW231123-like injection-recovery tests. A lensed \texttt{NRSur7dq4} signal with $m_g=0$, recovered with the same unlensed \texttt{IMRPhenomXPHM} template used in the real-data analysis, produces a similarly pronounced apparent nonzero graviton-mass posterior. Once lensing is included in the recovery, this spurious preference disappears. The similarity between the injection and real-data posteriors shows that the anomaly seen in GW231123 can be reproduced without any true modification of gravitational wave propagation.

The choice of \texttt{NRSur7dq4} for the injection is deliberate. Among the waveform families considered here, it provides the closest available surrogate-based proxy to a high fidelity numerical relativity reference. Following the LVK analyses of GW231123 \cite{LIGOScientific:2025cwb}, we compare \texttt{IMRPhenomXPHM}, \texttt{IMRPhenomXO4a}, and \texttt{NRSur7dq4}, which provide complementary probes of waveform systematic effects. Their convergence once lensing is included suggests that the dominant issue is not generic waveform disagreement, but the omission of point-mass lensing from the recovery model.

A preliminary noise-weighted mismatch study for GW231123-like parameters supports the same interpretation. In particular, the mismatch between a lensed \texttt{NRSur7dq4} reference signal and an unlensed \texttt{IMRPhenomXPHM} template is much larger, at the level of $\sim 1.5\times10^{-1}$, while the corresponding mismatch between lensed \texttt{NRSur7dq4} and lensed \texttt{IMRPhenomXPHM} is reduced to $\sim 2.4\times10^{-2}$. This shows that the dominant structural discrepancy is removed once the missing lensing physics is restored in the recovery model, and that the key issue is not merely few-percent waveform disagreement, but the much larger mismatch between a lensed signal and an unlensed recovery template.

These results identify a concrete failure mode in tests of gravity: for candidate point-mass lensed events, unmodeled wave-optics lensing can be absorbed by an unlensed waveform family and mimic a massive-graviton signal. At the same time, they strengthen the interpretation of GW231123 as a lensed candidate. Under the lensing-included interpretation, GW231123 is consistent with no evidence for a nonzero graviton mass and yields a single-event upper bound of $m_g < 4.3\times10^{-23}\,\mathrm{eV}/c^2$ at the 90\% credible level.

More broadly, the present results motivate extending this study beyond a single GW231123-like injection to determine more systematically when point-mass lensing distortions can be misidentified as propagation effects in gravitational wave analyses. In this sense, GW231123 is not only an interesting single event case study, but also a concrete starting point for lensing aware tests of gravity.

\noindent\textit{Acknowledgments---}
This work is supported by the National Natural Science Foundation of China Grants No. 12575063, and in part by ``the Special Funds for the Double First-Class Development of Wuhan University'' under the reference No. 2025-1302-010. Part of the numerical calculations in this paper have been done on the supercomputing system in the Supercomputing Center of Wuhan University. 

\bibliography{references}

\end{document}